\documentclass[a4paper]{article}

\usepackage{INTERSPEECH2020}

\usepackage{lipsum}
\usepackage{bm,amsmath,amssymb,mathrsfs,amsfonts}
\usepackage{caption}
\usepackage{scalefnt} %
\usepackage{multirow} %
\usepackage{enumitem}
\usepackage{type1cm}
\usepackage[noadjust]{cite}
\usepackage{url}
\usepackage{algorithm}
\usepackage{algpseudocode}
\usepackage{color}
\usepackage{threeparttable}
\usepackage{setspace}

\AtBeginDocument{
  \abovedisplayskip     =0.5\abovedisplayskip
  \abovedisplayshortskip=0.5\abovedisplayshortskip
  \belowdisplayskip     =0.5\belowdisplayskip
  \belowdisplayshortskip=0.5\belowdisplayshortskip}

\newcounter{num}

\algnewcommand{\Initialize}[1]{%
\State \textbf{Initialize:} {\raggedright #1}
}
\algnewcommand{\IIf}[1]{\State\algorithmicif\ #1\ \algorithmicthen}
\algnewcommand{\EndIIf}{\unskip\ \algorithmicend\ \algorithmicif}

\usepackage{etoolbox}

\newcommand{\algstrut}[1][\algruledefaultfactor]{\vrule width 0pt
depth .25\baselineskip height #1\baselineskip\relax}
\newcommand*{\algrule}[1][\algorithmicindent]{\hspace*{.5em}\vrule\algstrut
\hspace*{\dimexpr#1-.5em}}

\makeatletter
\newcount\ALG@printindent@tempcnta
\def\ALG@printindent{%
    \ifnum \theALG@nested>0%
    \ifx\ALG@text\ALG@x@notext%
    \else
    \unskip
    \ALG@printindent@tempcnta=1
    \loop
    \algrule[\csname ALG@ind@\the\ALG@printindent@tempcnta\endcsname]%
    \advance \ALG@printindent@tempcnta 1
    \ifnum \ALG@printindent@tempcnta<\numexpr\theALG@nested+1\relax%
    \repeat
    \fi
    \fi
}%

\patchcmd{\ALG@doentity}{\noindent\hskip\ALG@tlm}{\ALG@printindent}{}{\errmessage{failed to patch}}

\AtBeginEnvironment{algorithmic}{\lineskip0pt}

\title{Enhancing Monotonic Multihead Attention for Streaming ASR}
\name{Hirofumi Inaguma$^1$, Masato Mimura$^1$, Tatsuya Kawahara$^1$}
\address{
  $^1$Graduate School of Informatics, Kyoto University, Kyoto, Japan}
\email{\{inaguma,mimura,kawahara\}@sap.ist.i.kyoto-u.ac.jp}

\begin{document}

\maketitle
\vspace{-2mm}

\begin{abstract}
We investigate a monotonic multihead attention (MMA) by extending hard monotonic attention to Transformer-based automatic speech recognition (ASR) for online streaming applications.
For streaming inference, all monotonic attention (MA) heads should learn proper alignments because the next token is not generated until all heads detect the corresponding token boundaries.
However, we found not all MA heads learn alignments with a naïve implementation.
To encourage every head to learn alignments properly, we propose \textit{HeadDrop} regularization by masking out a part of heads stochastically during training.
Furthermore, we propose to prune redundant heads to improve consensus among heads for boundary detection and prevent delayed token generation caused by such heads.
Chunkwise attention on each MA head is extended to the multihead counterpart.
Finally, we propose \textit{head-synchronous} beam search decoding to guarantee stable streaming inference.
\end{abstract}
\noindent\textbf{Index Terms}: Transformer, streaming end-to-end ASR, monotonic multihead attention, beam search decoding

\vspace{-3mm}
\section{Introduction}
\vspace{-1.5mm}
Recent progress of end-to-end (E2E) automatic speech recognition (ASR) models bridges the gap from the state-of-the-art hybrid systems \cite{google_sota_asr}.
To make E2E models applicable to simultaneous interpretations in lecture and meeting domains, online streaming processing is necessary.
For E2E models, connectionist temporal classification (CTC) \cite{ctc_graves} and recurrent neural network transducer (RNN-T) \cite{rnn_transducer} have been dominant approaches and reached a level of real applications \cite{he2019streaming,sainath2020streaming}.
Meanwhile, attention-based encoder-decoder (AED) models \cite{chorowski2015attention,las} have demonstrated the powerful modeling capability in offline tasks \cite{s2s_comparison_google,s2s_comparison_baidu,rwth_end2end} and a number of streaming models have been investigated for RNN-based models \cite{hou2017gaussian,tjandra2017local,lawson2018learning,adaptive_computation_steps,moritz2019triggered_icassp2019,hard_monotonic_attention,mocha}.

Recently, the Transformer architecture \cite{vaswani2017attention}, based on self-attention and multihead attention, has shown to outperform the RNN counterparts in various domains \cite{karita2019comparative,zeyer2019comparison}, and several streaming models have been proposed such as triggered attention \cite{moritz2020streaming_icassp2020}, continuous-integrate-and-fire (CIF) \cite{cif}, hard monotonic attention (HMA) \cite{tsunoo2019towards,miao2020transformer}, and other variants \cite{tian2019synchronous}.
Triggered attention truncates encoder outputs by using CTC spikes and performs an attention mechanism over all past frames.
CIF learns acoustic boundaries explicitly and extracts context vectors from the segmented region.
Therefore, these models have adaptive segmentation policies relying on acoustic cues only.

On the other hand, HMA detects token boundaries on the decoder side by using lexical information as well.
Thus, it is more flexible for modeling non-monotonic alignments and has been investigated in simultaneous machine translation (MT) \cite{arivazhagan2019monotonic}.
Recently, HMA was extended to the Transformer architecture, named \textit{monotonic multihead attention} (MMA), by replacing each encoder-decoder attention head in the decoder with a monotonic attention (MA) head \cite{ma2019monotonic}.
Unlike a single MA head used in RNN-based models, each MA head can extract source contexts with different pace and learn complex alignments between input and output sequences.
Concurrently, similar methods have been investigated for Transformer-based streaming ASR \cite{tsunoo2019towards,miao2020transformer}.
Miao \textit{et al.} \cite{miao2020transformer} simplified the MMA framework by equipping a single MA head with each decoder layer to truncate encoder outputs as in triggered attention and perform attention over all past frames.
Tsunoo \textit{et al.} \cite{tsunoo2019towards} also investigated the MMA framework but resorted to using all past frames to obtain a decent performance.
However, looking back to the beginning of input frames lessens the advantage of linear-time decoding with HMA as the input length gets longer.

In this work, we investigate the MMA framework using restricted input context for the streaming ASR task.
To perform streaming recognition with the MMA framework, it is necessary for every MA head to learn alignments properly.
This is because the next token is not generated until all heads detect the corresponding token boundaries.
If some heads fail to detect the boundaries until seeing the encoder output of the final frame, the next token generation is delayed accordingly.
However, with a naïve implementation, we found that proper monotonic alignments are learnt in dominant MA heads only.
To prevent this, we propose \textit{HeadDrop}, in which a part of heads is entirely masked out at random as a regularization during training to encourage the rest non-masked heads to learn alignments properly.
Moreover, we propose to prune redundant MA heads in lower decoder layers to further improve consensus among heads on token boundary detection.
Chunkwise attention \cite{mocha} on top of each MA head is further extended to the multihead counterpart to extract useful representations and compensate the limited context size.
Finally, we propose \textit{head-synchronous} beam search decoding to guarantee streamable inference.

Experimental evaluations on Librispeech corpus show that our proposed methods effectively encourage MA heads to learn alignments properly, which leads to improvement of ASR performance.
Our optimal model enables stable streaming inference on other corpora as well without architecture modification.

\vspace{-3mm}
\section{Transformer ASR architecture}
\vspace{-1.5mm}
In this section, we detail the Transformer base architecture used in this paper.
Our Transformer architecture consists of stacked $E$ encoder layers followed by $C$ front-end CNN blocks and $D$ decoder layers \cite{karita2019comparative}.
A CNN block has two CNN layers with a $3 \times 3$ filter with a channel size $32$ followed by a ReLU activation.
The frame rate is reduced by a factor of $2^{C}$ by a max-pooling layer with a stride of $2 \times 2$ after every block.
Each encoder layer is composed of a self-attention (SAN) sub-layer followed by a position-wise feed-forward network (FFN) sub-layer, wrapped by residual connections and layer normalization \cite{ba2016layer}.
A key component of SAN sub-layers is a multihead attention (MHA) mechanism, in which key, value, and query matrices are split into $H$ potions with a dimension $d_{k}=d_{\rm model}/H$ after linear transformations and each head performs a scaled-dot attention mechanism: $\mbox{Attention}(Q,K,V) = \mbox{softmax}(QK^{\mathsf T}/\sqrt{d_{k}})V$, where $K$, $Q$, and $V$ represent key, query, and value matrices on each head, respectively.
The outputs from all heads are concatenated along the feature dimension followed by a linear transformation.
A FFN sub-layer is composed of two linear layers with the inner dimension $d_{\rm ff}$, interleaved with a ReLU activation between them.

In each decoder layer, unlike the encoder layer, additional encoder-decoder attention sub-layer is inserted between SAN and FFN sub-layers, and causal masking is performed to prevent the decoder from seeing the future tokens.
We adopt three 1D-convolutional layers for positional embeddings \cite{mohamed2019transformers}.
The entire network is optimized by minimizing the negative log-likelihood and CTC loss with an interpolation weight $\lambda_{\rm ctc}=0.3$ \cite{karita2019comparative}.

\vspace{-3mm}
\section{Monotonic multihead attention (MMA)}\label{sec:mma}
\vspace{-1.5mm}
In this section, we review hard monotonic attention (HMA) \cite{hard_monotonic_attention}, monotonic chunkwise attention (MoChA) \cite{mocha}, and monotonic multihead attention (MMA) \cite{ma2019monotonic} as an extension of them.

\vspace{-2.5mm}
\subsection{Hard monotonic attention (HMA)}\label{ssec:hard_monotonic_attention}
\vspace{-2mm}
HMA was originally proposed for online linear-time decoding with RNN-based AED models.
At output step $i$, the decoder scans encoder outputs from left to right and stops at an index $j=t_{i}$ (token boundary) to attend the corresponding single encoder output $h_{j}$.
The decoder has options to stop at the current index or move forward to the next index.
The next boundary is determined by resuming scanning from the previous boundary.
As hard attention is not differentiable, the expected alignments $\alpha_{i,j}$ are calculated by marginalizing over all possible paths during training as follows:
\begin{eqnarray}
\alpha_{i,j} &=& p_{i,j}\bigg((1-p_{i,j-1})\frac{\alpha_{i,j-1}}{p_{i,j-1}}+\alpha_{i-1,j}\bigg) \label{eq:monotonic_attention_alpha} \\
p_{i,j} &=& \mbox{Sigmoid}(\mbox{MonotonicEnergy}(h_{j}, s_{i}))
\label{eq:monotonic_energy}
\end{eqnarray}
where $p_{i,j}$ is a selection probability and a monotonic energy function $\mbox{MonotonicEnergy}(\cdot)$ takes the $i$-th decoder state $s_{i}$ and $j$-th encoder output $h_{j}$ as inputs.
Whenever $p_{i,j} \geq 0.5$ is satisfied at test time, $\alpha_{i,j}$ is activated (i.e., set to 1.0).

\vspace{-2.5mm}
\subsection{Monotonic chunkwise attention (MoChA)}\label{ssec:mocha}
\vspace{-2mm}
To relax strict input-output alignment by using the surrounding contexts, MoChA introduces additional soft attention mechanism on top of HMA.
Given the boundary $j$, chunkwise attention is performed over a fixed window of width $w$ from there:
\begin{gather}
\beta_{i,j} = \sum_{k=j}^{j+w-1}\bigg(\alpha_{i,k}{\rm exp}(u_{i,j})/\sum_{l=k-w+1}^{k}{{\rm exp}(u_{i,l})}\bigg) \label{eq:mocha_beta}
\end{gather}
where $u_{i,j}$ is a chunk energy parameterized similar to the monotonic energy in Eq. \eqref{eq:monotonic_energy} using separate parameters.
$\alpha_{i,j}$ in Eq. \eqref{eq:mocha_beta} is a continuous value during training, but is a binary value according to $p_{i,j}$ at test time.

\vspace{-2.5mm}
\subsection{Monotonic multihead attention (MMA)}\label{ssec:mma}
\vspace{-2mm}
To keep the expressive power of Transformer with the multihead attention mechanism while enabling online linear-time decoding, MMA was proposed as an extension of HMA \cite{ma2019monotonic}.
Each encoder-decoder attention head in the decoder is replaced with a monotonic attention (MA) head in Eq. \eqref{eq:monotonic_attention_alpha} by defining the monotonic energy function in Eq. \eqref{eq:monotonic_energy} as follows:
\begin{gather}
\mbox{MonotonicEnergy}(h_{j}, s_{i}) = \frac{W_{s}s_{i}(W_{h}h_{j})^{\mathsf T}}{\sqrt{d_{k}}} + r \label{eq:monotonic_energy_transformer}
\end{gather}
where $W_{s}$ and $W_{h}$ are parameter matrices, and $r$ is a learnable offset parameter (initialized with $-2$ in this work).
Unlike a case of a single MA head in Section \ref{ssec:hard_monotonic_attention}, each MA head can attend to input speech with different pace because its decision process regarding timing to activate $\alpha_{i,j}$ does not influence each other at each output step.
The side effect is that all heads must be activated to generate a token.
Otherwise, some MA heads continue to scan encoder outputs until the last time index, which leads to significant increase of latency during inference.

Furthermore, unlike previous works \cite{ma2019monotonic,tsunoo2019towards} having a single chunkwise attention (CA) head on each MA head, we extend it to the multi-head version having $H_{\rm ca}$ heads per MA head to extract useful representations with multiple views from each boundary (\textit{chunkwise multihead attention}).
Assuming each decoder layer has $H_{\rm ma}$ MA heads, the total number of CA heads is $H_{\rm ma} \cdot H_{\rm ca}$ at the layer.
However, we found that sharing parameters of CA heads among MA heads in the same layer is effective in our pilot experiments and adopt this strategy in all experiments.
The chunk energy $u_{i,j}$ for each CA head is designed as in Eq. \eqref{eq:monotonic_energy_transformer} without $r$.

\begin{figure}[t]
  \centering
  \vspace{-3mm}
  \includegraphics[width=0.90\linewidth]{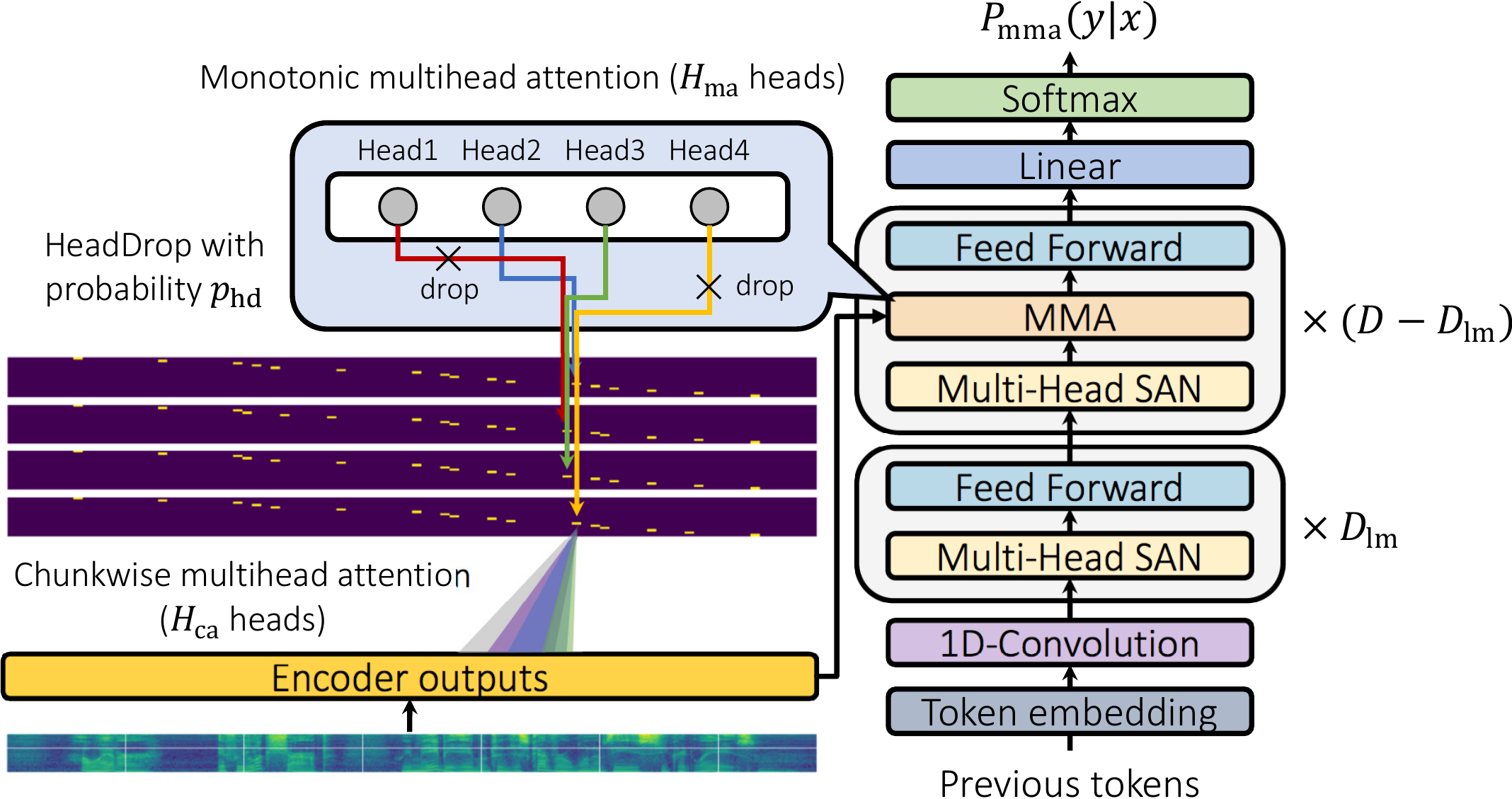}
  \vspace{-2.5mm}
  \caption{System overview. Residual connections and layer normalization are omitted.}
  \label{fig:system_overview}
  \vspace{-7mm}
\end{figure}

\vspace{-3mm}
\section{Enhancing monotonic alignments}\label{sec:enhance}
\vspace{-1.5mm}
In the Transformer models, there exist many attention heads and residual connections, so it is unclear that all heads contribute to the final predictions.
Michel \textit{et al.} \cite{michel2019sixteen} reported that most heads can be pruned at test time without significant performance degradation in standard MT and BERT \cite{devlin2018bert} architectures.
They also revealed that important heads are determined in early training stage.
Concurrently, Voita \textit{et al.} \cite{voita2019analyzing} also reported the similar observations by automatically pruning a part of heads with $L_{0}$ penalty \cite{louizos2017learning}.
In our preliminary experiments, we also observed that not all MA heads learn alignments properly in the MMA-based ASR models and monotonic alignments are learnt only by dominant heads in upper decoder layers.
Since $\alpha_{i,j}$ in Eq. \eqref{eq:monotonic_attention_alpha} are not normalized over inputs so as to sum up to one during training, context vectors from heads which do not learn alignments are more likely to become zero vectors at test time.
This is a severe problem because (1) it leads to mismatch between training and testing conditions, and (2) the subsequent tokens cannot be generated until all heads are activated.
To alleviate this problem, we propose the following methods.

\vspace{-3.3mm}
\subsection{HeadDrop}\label{ssec:headdrop}
\vspace{-2mm}
We first propose a regularization method to encourage each MA head to equally contribute to the target task.
During training, we stochastically zero out all elements in each MA head (i.e., $\alpha_{i,j}=0, 1 \leq j^{\forall} \leq |\bm{h}|$) with a fixed probability $p_{\rm hd}$ to force the other heads to learn alignments.
The decision of dropping each head is independent of other heads regardless of the depth of the decoder.
The output of a MMA function is normalized by dividing it by $H_{\rm ma}^{+}/H_{\rm ma}$, where $H_{\rm ma}^{+}$ is the number of non-masked MA heads.
We name this \textit{HeadDrop}, inspired by dropout \cite{dropout} and DropConnect \cite{wan2013regularization}.

\vspace{-3mm}
\subsection{Pruning monotonic attention heads in lower layers}\label{ssec:head_pruning}
\vspace{-2mm}
Although HeadDrop is effective for improving the contribution of each MA head, we found that some MA heads in the lower decoder layers still do not learn alignments properly.
Therefore, we propose to prune such redundant heads because they are harmful for streaming decoding.
We remove the MMA function in the first $D_{\rm lm}$ decoder layers from the bottom ($0 \leq D_{\rm lm} \leq D-1$) during both training and test time (see Figure \ref{fig:system_overview}).
These layers have SAN and FFN sub-layers only and serve as a pseudo language model (LM).
The total number of effective MA heads $H^{\rm total}_{\rm ma}$ is $(D - D_{\rm lm}) \cdot H_{\rm ma}$.
This method is also based on findings in \cite{voita2019analyzing} that the lower layers of the Transformer decoder are mostly responsible for language modeling.
Another advantage of pruning redundant heads is that inference speed is improved, which is effective for streaming ASR.

\vspace{-3.4mm}
\section{Head-synchronous beam search decoding}\label{sec:head_sync}
\vspace{-1.5mm}
During beam search in the MMA framework, failure of boundary detection in some MA heads in some beam candidates at an output step easily prevents the decoder from continuing streaming inference.
This is because other candidates must wait for the hypothesis pruning until all heads in all candidates are activated at each output step.
To continue streaming inference, we propose \textit{head-synchronous} beam search decoding (Algorithm \ref{algo:head_sync}).
The idea is to force non-activated heads to activate after a small fixed delay.
If a head at the $l$-th layer cannot detect a boundary for $\epsilon_{\rm wait}$ frames after the leftmost boundary detected by other heads in the same layer, the boundary of such a head is set to the rightmost boundary $t_{\rm tail}^{l}$ among already detected boundaries at the current output step (line \ref{algo:force_activate}).
Therefore, latency between the fastest (rightmost) and slowest (leftmost) boundary positions in the same layer is less than $\epsilon_{\rm wait}$ frames.
We note that the decisions of boundary detection at the $l$-th layer are dependent on outputs from the $(l-1)$-th layer, and at least one head must be activated at each layer to generate a token.
For the efficient implementation, we search boundaries of all heads in a layer in parallel, thus the loop in line \ref{algo:head_loop} can be ignored.
Moreover, we perform batch processing over multiple hypotheses in the beam and cache previous decoder states for efficiency.
Note that head synchronization is not performed during training to maintain the divergence of boundary positions.
Thus, synchronization can have the ensemble effect for boundary detection.

\vspace{-3.4mm}
\section{Experimental evaluations}\label{sec:experiments}
\vspace{-1.5mm}
\subsection{Experimental setup}\label{ssec:setup}
\vspace{-1.5mm}
We used the 960-hour Librispeech dataset \cite{librispeech} for experimental evaluations.
We extracted 80-channel log-mel filterbank coefficients computed with a 25-ms window size shifted every 10 ms using Kaldi \cite{kaldi}.
We used a 10k vocabulary based on the Byte Pair Encoding (BPE) algorithm \cite{sennrich2015neural}.
For the Transformer model configurations, we used ($d_{\rm model}$, $d_{\rm ff}$, $H$, $H_{\rm ma}$, $H_{\rm ca}$, $w$, $C$, $E$, $D$) $=$ ($256$, $2048$, $4$, $4$, $1$, $4$, $3$, $12$, $6$) for the baseline MMA models.
Adam \cite{adam} optimizer was used for training with Noam learning rate schedule \cite{vaswani2017attention}.
Warmup steps and a learning rate constant were set to $25000$ and $5.0$, respectively.
We averaged model parameters at the last $10$ epochs for evaluation.
Both dropout and label smoothing \cite{label_smoothing} were used with a probability $0.1$.
We set $p_{\rm hd}$ to $0.5$.
We used a 4-layer LSTM LM with $1024$ memory cells for decoding with a beam width of $10$.
We used decoding hyperparameters ($\alpha$, $\beta$) $=$ ($0.5$, $2.0$).\footnote{Code: \scriptsize{\url{https://github.com/hirofumi0810/neural_sp}}.}

\newcommand{\pij}{p_{i,j}^{l,h}}
\newcommand{\ti}{t_{i}^{l,h}}
\newcommand{\ttail}{t_{\rm tail}^{l}}

\begin{algorithm}[t]
\caption{Head-synchronous beam search decoding}
\label{algo:head_sync}
\begin{algorithmic}[1]
    \scriptsize
    \Require{${\bm h}$: encoder outputs, $\epsilon_{\rm wait}$: wait time threshold, $B$: beam width}
    \Ensure{$\Omega_{\rm end}$}: top-k hypotheses
    \Initialize{$t_{0}^{l,h} \gets 1$, $\Omega \gets \{\}$, $\Omega_{\rm end} \gets \{\}$, $L_{\rm max} \gets 200$}

    \For {$i \leftarrow 1$ to $L_{\rm max}$}
        \State $\Omega_{\rm new} \gets \{\}$
        \For {$k \leftarrow 1$ to $|\Omega|$} \Comment{Batchfy}
            \For {$l$ $\leftarrow$ $D_{\rm lm}$+1 to $D$}
                \State $\ttail \gets 1$
                \For {$h$ $\leftarrow$ 1 to $H_{\rm ma}$ in parallel} \label{algo:head_loop}
                    \State $\pij=\mbox{Sigmoid}(\mbox{MonotonicEnergy}(h_{j},s_{i-1}^{l-1}))$
                    \For {$j$ $\leftarrow$ $t_{i-1}^{l,h}$ to $|{\bm h}|$}
                        \If {$\pij \geq 0.5$}
                            \State $\ti \leftarrow j$;\hspace{1mm}$\ttail \gets \max(\ttail,\ j)$;\hspace{1mm}\textbf{break};
                        \Else
                            \If {\color{blue}{$j \geq \ttail + \epsilon_{\rm wait}$}\color{black}}
                                \State \color{blue}{$\ti \leftarrow \ttail$}\color{black} \label{algo:force_activate} \Comment{Forced activation}
                                \State \textbf{break}
                            \EndIf
                        \EndIf
                     \EndFor
                \EndFor
            \EndFor
            \State Append $\mbox{topK}(\log{P_{\rm mma}} + \alpha \log{P_{\rm lm}} + \beta i)$ to $\Omega_{\rm new}$
        \EndFor
        \State $\Omega,\ \Omega_{\rm end} \gets \mbox{Prune}(\Omega_{\rm new},\ \Omega_{\rm end})$
        \IIf {$i = L_{\rm max}$ or $|\Omega_{\rm end}| = B$} \textbf{break};
        \EndIIf
    \EndFor
    \State \Return $\Omega_{\rm end}$
\end{algorithmic}
\end{algorithm}
\setlength{\textfloatsep}{2.0mm}  %

\vspace{-3mm}
\subsection{Evaluation measure of boundary detection}\label{sec:evaluation_metric}
\vspace{-2mm}
To assess consensus among heads for token boundary detection, we propose metrics to evaluate (1) how well each MA head learns alignments ({\bf \textit{boundary coverage}}) and (2) how often the model satisfies the streamable condition ({\bf \textit{streamability}}).
This is because even if better word error rate (WER) is obtained, the model cannot continue streaming inference if some heads do not learn alignments well, whose evaluation is missing in \cite{miao2020transformer,tsunoo2019towards}.

\vspace{-3mm}
\subsubsection{Boundary coverage}\label{ssec:attention_coverage}
\vspace{-2mm}
During beam search, we count the total number of boundaries ($j$ such that $\alpha_{:,j}=1$) up to the $i$-th output step averaged over all MA heads, $Q^{n,k}_{i}$, for every candidate $\tilde{y}^{n,k}$ in the $n$-th utterance:
\begin{gather*}
Q^{n,k}_{i} = \frac{1}{H_{\rm ma}^{\rm total}}\sum_{h=1}^{H_{\rm ma}^{\rm total}}\sum_{i'=1}^{i}\sum_{j=1}^{|\bm{h}|}\alpha_{i',j}^{h}
\end{gather*}
The boundary coverage $R_{\rm cov}$ is defined as the ratio of $Q^{n,1}_{|\tilde{y}^{n,1}|}$ to the corresponding hypothesis length $|\tilde{y}^{n,1}|$ of the best candidate and averaged over $N$ utterances in the evaluation set as follows:
\begin{gather*}
R_{\rm cov} \ [\%] = \frac{1}{N}\sum_{n=1}^{N}\frac{Q^{n,1}_{|\tilde{y}^{n,1}|}}{|\tilde{y}^{n,1}|} \times 100
\end{gather*}

\vspace{-3.5mm}
\subsubsection{Streamability}\label{ssec:streamability}
\vspace{-2mm}
The streamability $R_{\rm str}$ is defined as the ratio of utterances satisfying a condition where $Q^{n,k}_{i}=|\tilde{y}^{n,k}|$ over all candidates up to the $|\tilde{y}^{n,1}|$-th output step (i.e, until generation of the best hypothesis is completed) as follows:
\begin{gather*}
R_{\rm str} \ [\%] = \frac{1}{N}\sum_{n=1}^{N}\delta_{n} \times 100 \\
\delta_{n} = \begin{cases}
    1 & (Q^{n,k}_{i} = |\tilde{y}^{n,k}|, \ 1 \leq i^{\forall} \leq |\tilde{y}^{n,1}|, 1 \leq k^{\forall} \leq |\Omega^{n}_{i}|) \\
    0 & (otherwise)
\end{cases}
\end{gather*}
where $\delta_{n}$ is the delta function and $\Omega^{n}_{i}$ is a hypothesis set at $i$-th output step of the $n$-th utterance.
$\delta_{n}=0$ indicates that the model failed streaming recognition somewhere in the $n$-th utterance, i.e., continued scanning the encoder outputs until the last frame.
However, we note that it does not mean the model leverages additional context.

\vspace{-3mm}
\subsection{Offline ASR results}\label{ssec:result_offline}
\vspace{-2mm}
\noindent\textbf{{HeadDrop and pruning MA heads in lower layers improve WER and streamability}}\hspace{2mm}
Table \ref{tab:result_librispeech_baseline} shows the results for offline MMA models on the Librispeech dev-clean/other sets.
"Offline" means the encoder is the offline architecture.
Boundary coverage and stremability were averaged over two sets.
A naïve implementation {\tt A1} showed a very poor performance.
By pruning MA heads in lower layers with increasing $D_{\rm lm}$, WER was
significantly reduced, but the boundary coverage was not so high ({\tt A5}).
The proposed HeadDrop also significantly improved WER, and with the increase of $D_{\rm lm}$, the boundary coverage was drastically improved to almost 100\% ({\tt B3}-{\tt B5}).
We can conclude that MA heads in lower layers are not necessarily important.
This is probably because (1) the modalities between input and output sequences are different in the ASR task and (2) hidden states in lower decoder layers tend to represent the previous tokens, thus are not suitable for alignment.

\begin{table}[t]
    \centering
    \scriptsize
    \begingroup
    \vspace{-2.5mm}
    \caption{Results of \underline{offline} MMA models on the \underline{dev sets} with \textbf{standard} beam search decoding. HD: HeadDrop.}\label{tab:result_librispeech_baseline}
    \vspace{-3mm}
    \begin{tabular}{c|c|c|c|c|ccc} \toprule
     \multirow{2}{*}{ID} & \multirow{2}{*}{$D_{\rm lm}$} & \multirow{2}{*}{$H_{\rm ma}$} & \multirow{2}{*}{$H_{\rm ma}^{\rm total}$} & \multirow{2}{*}{HD} & \multicolumn{3}{c}{dev-clean / dev-other} \\ \cline{6-8}
      &  &  &  &  & \%WER & $R_{\rm cov}$ & $R_{\rm str}$ \\ \hline
      {\tt A1} & 0 & \multirow{5}{*}{4} & 24 & \multirow{5}{*}{-} & 8.6 / 16.5 & 67.40 & 0.0 \\
      {\tt A2} & 1 &  & 20 &  & 7.3 / 16.3 & 79.02 & 0.0 \\
      {\tt A3} & 2 &  & 16 &  & 4.7 / 12.6 & 86.07 & 0.0 \\
      {\tt A4} & 3 &  & 12 &  & 4.5 / 12.8 & 83.87 & 0.0 \\
      {\tt A5} & 4 &  & 8 &  & {\bf 3.6} / {\bf 10.8} & {\bf 93.80} & {\bf 0.9} \\
    \hline

      {\tt B1} & 0 & \multirow{5}{*}{4} & 24 & \multirow{5}{*}{\checkmark} & {\bf 3.7} / 11.4 & 60.59 & 0.0 \\
      {\tt B2} & 1 &  & 20 &  & 4.0 / 11.9 & 73.73 & 0.0 \\
      {\tt B3} & 2 &  & 16 &  & 3.9 / {\bf 10.8} & 98.85 & 3.7 \\
      {\tt B4} & 3 &  & 12 &  & 4.1 / 11.0 & 99.36 & 6.4 \\
      {\tt B5} & 4 &  & 8 &  & 4.1 / 11.3 & {\bf 99.50} & {\bf 15.8} \\
    \hline

      {\tt C1} & \multirow{3}{*}{0} & 1 & 6 & - & 4.9 / 11.7 & 99.38 & 15.7 \\
      {\tt C2} &  & 1 & 6 & \checkmark & 3.7 / {\bf 10.4} & {\bf 99.86} & {\bf 35.9} \\
      {\tt C3} &  & 2 & 12 & \checkmark & {\bf 3.5} / 10.7 & 72.08 & 0.0 \\
       \bottomrule
    \end{tabular}
    \vspace{-3.5mm}
    \endgroup
\end{table}

\begin{table}[t]
    \centering
    \scriptsize
    \begingroup
    \caption{Results of \underline{offline} MMA models on the \underline{dev sets} with \textbf{head-synchronous} beam search decoding and HeadDrop.}\label{tab:result_librispeech_head_sync}
    \vspace{-3mm}
    \begin{tabular}{c|c|c|c|c|ccc} \toprule
     \multirow{2}{*}{ID} & \multirow{2}{*}{$D_{\rm lm}$} & \multirow{2}{*}{$H_{\rm ma}$} & \multirow{2}{*}{$w$} & \multirow{2}{*}{$H_{\rm ca}$} & \multicolumn{3}{c}{dev-clean / dev-other} \\ \cline{6-8}
      &  &  &  &  & \%WER & $R_{\rm cov}$ & $R_{\rm str}$ \\ \hline
      {\tt B3} & 2 & \multirow{3}{*}{4} & \multirow{3}{*}{4} & \multirow{3}{*}{1} & 3.9 / 10.7 & 99.74 & 21.6 \\
      {\tt B4} & 3 &  &  &  & 3.9 / {\bf 10.6} & {\bf 99.76} & 25.1 \\
      {\tt B5} & 4 &  &  &  & {\bf 3.8} / 11.1 & 99.84 & {\bf 40.5} \\
      \hline

      {\tt D1} & 2 & \multirow{3}{*}{4} & \multirow{3}{*}{16} & \multirow{3}{*}{1} & {\bf 3.3} / {\bf 9.9} & 99.78 & 37.4 \\
      {\tt D2} & 3 &  &  &  & 3.7 / 10.8 & 99.83 & 36.5 \\
      {\tt D3} & 4 &  &  &  & 3.5 / 10.4 & {\bf 99.93} & {\bf 60.4} \\
      \hline

      {\tt E1} & 2 & \multirow{6}{*}{4} & \multirow{6}{*}{16} & \multirow{3}{*}{2} & {\bf 3.3} / 10.2 & 99.78 & 40.6 \\
      {\tt E2} & 3 &  &  &  & 3.6 / 10.3 & 99.87 & 51.2 \\
      {\tt E3} & 4 &  &  &  & 3.5 / 10.7 & {\bf 99.92} & 50.0 \\ \cline{2-2} \cline{5-5}
      {\tt E4} & 2 &  &  & \multirow{3}{*}{4} & {\bf 3.3} / {\bf 9.8} & 99.91 & 77.9 \\
      {\tt E5} & 3 &  &  &  & 3.4 / 9.9 & 99.90 & {\bf 84.5} \\
      {\tt E6} & 4 &  &  &  & 3.6 / 10.4 & {\bf 99.92} & 63.2 \\ \hline
      
      {\tt F1} & 0 & 1 & 16 & 4 & 3.5 / 10.5 & 96.23 & 40.6 \\
      
       \bottomrule
    \end{tabular}
    \endgroup
    \vspace{-1mm}
\end{table}

\begin{table}[t]
    \centering
    \scriptsize
    \begingroup
    \vspace{-2.5mm}
    \caption{Results of \underline{streaming} MMA models on the Librispeech, TEDLIUM2, and AISHELL-1 \underline{test sets}}\label{tab:result_final}
    \vspace{-3mm}
    \begin{tabular}{c|l|cc|c|c} \toprule
     \multicolumn{2}{c|}{\multirow{3}{*}{Model}} & \multicolumn{3}{c|}{\%WER} & \%CER \\ \cline{3-6}
     \multicolumn{2}{c|}{} & \multicolumn{2}{c|}{Librispeech} & \multirow{2}{*}{\shortstack{TED\\LIUM2}} & \multirow{2}{*}{\shortstack{AISH\\ELL-1}} \\ \cline{3-4}
     \multicolumn{2}{c|}{} & clean & other &  &  \\ \hline

     \multirow{4}{*}{\rotatebox{90}{\scriptsize{Offline}}}
         & Transformer (ours) & 3.3 & 9.1 & 10.1 & 6.4 \\
         & + data augmentation & 2.8 & 7.0 & - & - \\
         & ++ large model & 2.5 & 6.1 & - & - \\ \cline{2-6} %

         & MMA ({\tt E5}) & 3.4 & 9.9 & 10.5 & 6.5 \\
         \hline

     \multirow{8}{*}{\rotatebox{90}{\scriptsize{Streaming}}}
         & Triggered attention \cite{moritz2020streaming} & 2.8 & 7.2 & - & - \\
         & CIF \cite{cif} & 3.3 & 9.7 & - & - \\
         & MoChA \cite{ctc_sync} & 4.0 & 9.5 & 11.3 & - \\
         & MMA ($w=\infty$) \cite{tsunoo2019towards} & - & - & - & 9.7 \\ \cline{2-6}
         & MMA (narrow chunk) & 3.5 & 11.1 & 11.0 & 7.5 \\
         & MMA (wide chunk) & 3.3 & 10.5 & {\bf 10.2} & {\bf 6.6} \\
         & + data augmentation & 3.0 & 8.5 & - & - \\
         & ++ large model & {\bf 2.7} & {\bf 7.1} & - & - \\
       \bottomrule
    \end{tabular}
    \endgroup
    \vspace{-1mm}
\end{table}

\vspace{0.3mm}
\noindent\textbf{{Head-synchronous beam search decoding improves WER and streamability}}\hspace{2mm}
Next, the results with head-synchronous beam search decoding are shown in Table \ref{tab:result_librispeech_head_sync}.
$\epsilon_{\rm wait}$ is set to 8 in all models.
Head-synchronous decoding improved both boundary coverage and streamability.
We found that if a head cannot detect the boundary around the corresponding actual acoustic boundary, it tends to stop around the next acoustic boundary twice to compensate the failure when using a standard beam search.
Head-synchronous decoding alleviated this mismatch of boundary positions and led to small WER improvement.

\vspace{0.3mm}
\noindent\textbf{{Chunkwise multihead attention is effective}}\hspace{2mm}
Furthermore, we increased the window size $w$ and number of heads $H_{\rm ca}$ in chunkwise attention, both of which further improved WER.
With $D_{\rm lm}=4$, {\tt E3} and {\tt E6} did not obtain benefits from larger $H_{\rm ca}$.
Increasing $w$ to longer than 16 was not effective.

\vspace{0.3mm}
\noindent\textbf{{Multiple MA heads in each layer are necessary}}\hspace{2mm}
We also examined the effect of the number of MA heads $H_{\rm ma}$ in each layer ({\tt C1}-{\tt C3} in Table \ref{tab:result_librispeech_baseline}).
{\tt C1} with only one MA head per layer showed a high boundary coverage and was further improved with HeadDrop ({\tt C2}).
{\tt C3} with two heads per layer degraded streamability very much.
Although {\tt C2} showed better performances than {\tt B*}, it did not obtain much gains from larger $w$ and $H_{\rm ma}$ ({\tt F1} in Table \ref{tab:result_librispeech_head_sync}).
This confirms that having multiple MA heads in upper layers is more effective than simply reducing the number of MA heads per layer.
In other words, the place of MA heads is more important than the total number of them.

Here, what does the rest $15.5$\% for streamability in {\tt E5} account for?
We found that the last few tokens corresponding to the tail part of input frames were predicted after head pointers on upper layers reached the last encoder output.
For these $15.5$\% utterances, {\tt E5} was able to continue streaming decoding until $76.9$\% of the input frames on average.
Since the tail part is mostly silence, this does not affect streaming recognition.
In our manual analysis, we observed that MA heads in the same layer move forward with the similar pace, and the pace gets faster in upper layers.\footnote{Examples available at \scriptsize{\url{https://github.com/hirofumi0810.github.io/demo/enhancing_mma_asr}}.}
This is because decoder states are dependent on the output from the lower layer.
Considering the balance between WER and streamability performance, we will use the {\tt E5} setting for streaming experiments in the next section.

\vspace{-3mm}
\subsection{Streaming ASR results}\label{ssec:result_streaming}
\vspace{-2mm}
Finally, we present the results of streaming MMA models for the Librispeech test sets in Table \ref{tab:result_final}.
We also included results on TEDLIUM2 \cite{tedlium2} and AISHELL-1 \cite{aishell} to confirm whether the optimal configuration tuned on Librispeech can work in other corpora as well.
We adopted the chunk hopping mechanism \cite{dong2019self} for the online encoder.
Following \cite{cif,miao2020transformer}, we set the left/current (hop)/right chunk sizes to $960/640/320$ (narrow chunk) and $640/1280/640$ (wide chunk) [ms].
We used speed perturbation \cite{speed_perturbation} and SpecAugment \cite{specaugment} for data augmentation, but speed perturbation was applied by default for TEDLIUM2 and AISHELL-1.
For large models, we used ($d_{\rm model}$, $d_{\rm ff}$, $H$) $=$ ($768$, $3072$, $8$) and other hyperparameters were kept.
Head-synchronous decoding was used for all MMA models.
We used CTC scores during inference only for standard Transformer models.
Our streaming MMA models achieved comparable results to the state-of-the-art Transformer-based streaming ASR model \cite{moritz2020streaming} without looking back to the first input frame.
Moreover, our model outperformed the MMA model with $w=\infty$ \cite{tsunoo2019towards} by a large margin.
Increasing the model size was also effective.
The streamabilities of the streaming MMA models on TEDLIUM2 and AISHELL-1 with the wide chunk were 80.0\%, and 81.5\%, respectively.
This confirms that the {\tt E5} setting generalizes to other corpora without architecture modification.

\vspace{-3mm}
\section{Conclusion}\label{sec:conclusion}
\vspace{-1.5mm}
We tackled the alignment issue in monotonic multihead attention (MMA) for online streaming ASR with \textit{HeadDrop} regularization and head pruning in lower decoder layers.
We also stabilized streamable inference by \textit{head-synchronous} decoding.
Our future work includes investigation of adaptive policies for head pruning and regularization methods to make the most of the MA heads instead of discarding them.
Minimum latency training as done in MoChA \cite{inaguma2020streaming} is another interesting direction.

\tiny
\bibliographystyle{IEEEtran}
\bibliography{reference}

\end{document}